# Fuel Pellet Alignment in Heavy-Ion Inertial Fusion Reactor

T. Kubo[1], T. Karino[1], H. Kato[1] and S. Kawata[1]
[1]Graduate School of Engineering, Utsunomiya University, Japan

*Abstract*—In inertial confinement fusion, the scientific issues include the generation and transport of driver energy, the pellet design, the uniform target implosion physics, the realistic nuclear fusion reactor design, etc. In this paper, we present a pellet injection into a power reactor in heavy ion inertial fusion. We employ a magnetic correction method to reduce the pellet alignment error in heavy ion inertial fusion reactor chamber, including the gravity, the reactor gas drag force and the injection errors. We found that the magnetic correction device proposed in this paper is effective to construct a robust pellet injection system with a sufficiently small pellet alignment error.

*Index Terms*— inertial confinement fusion, pellet injection, pellet alignment error, heavy ion inertial fusion.

## I. Introduction

In inertial confinement fusion (ICF) the issues to be studied include the generation and transport of the driver energy to a fuel target, the optimum pellet structure, the realistic nuclear fusion reactor design, the target alignment and injection in a reactor, the degradation of fusion energy output due to the non-uniformity target implosion, etc. [1-13]. In this paper, we present a new method of the fuel-target alignment and injection, and propose a magnetic correction method [14] to reduce the target alignment error actively in a heavy ion beam (HIB) ICF (HIF) reactor.

The heavy ion beam (HIB) fusion (HIF) has been proposed in 1970's [15]. The HIF reactor designs were also proposed in Refs. [16-18]. HIB ions deposit their energy inside of materials, and the interaction of the HIB ions with the materials are well understood [1, 19]. The HIB ion interaction with a material is explained and defined well by the Coulomb collision and a plasma wave excitation in the material plasma. The HIB ions deposit all the HIB ion energy inside of the material. The HIB energy deposition length is typically the order of ~0.5mm in a HIF fuel target depending on the HIB ion energy and the material. When several MJ of the HIB energy is deposited in the material in a fuel target, the temperature of the energy deposition layer plasma becomes about 300 eV or so. The peak temperature or the peak plasma pressure appears near the HIB ion stopping area by the Bragg peak effect, which comes from the nature of the Coulomb collision [19]. The total stopping range would be normally wide inside of the solid material. The relatively large density gradient scale length is created in the HIBs energy deposition region in an DT fuel target, and it may also contribute to reduce the R-T instability growth rate [20, 21].

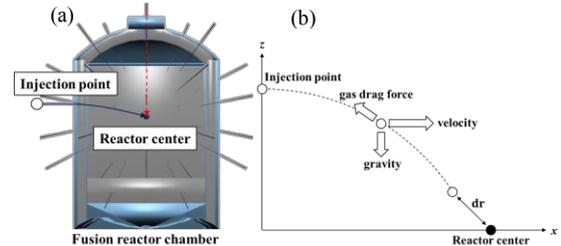

Fig. 1. (a) A fusion reactor chamber. A fuel pellet is injected from the outside of the reactor to the reactor center. In the reactor chamber, a reactor gas, for instance He, would be filled, and the fuel pellet would receive the gas drag force and the gravity (see Fig. 1(b)).

In HIF reactor system. HIB driver accelerators have a high driver energy efficiency of 30-40 % from the electricity to the HIB energy. The high driver efficiency would relax the requirement for the fuel target gain. In HIF the target gain of 50~70 allows us to construct HIF fusion reactor systems, and 1GW of the electricity output would be realized. The HIB accelerator has a high controllability to define the ion energy,

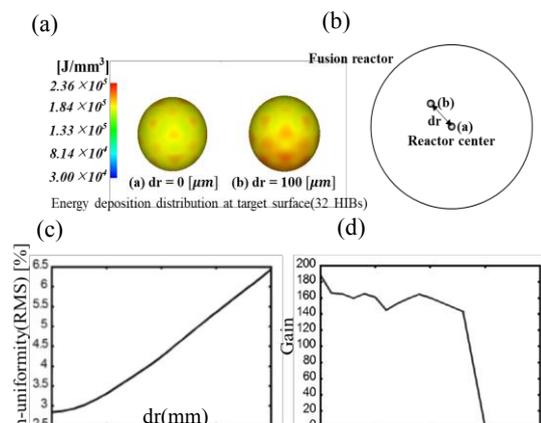

Fig. 2. When the HIBs (heavy ion beams) introduce an illumination non-uniformity as shown in (a) due to the target alignment error (b) *dr*, the fuel pellet implosion non-uniformity is induced as presented in (c). The example simulation results [1] are shown in (c) for the HIBs illumination non-uniformity versus *dr* and in (d) for the target gain versus *dr*.

The work was partly supported by JSPS, MEXT, U.S.-Japan exchange program and CORE (Center for Optical Research and Education, Utsunomiya University).

T. Kubo, S. Kondo, T. Karino, H. Kato and S. Kawata is with Graduate School of Engineering, Utsunomiya University, Utsunomiya 321-8585, Japan (e-mail: kwt@cc.utsunomiya-u.ac.jp).

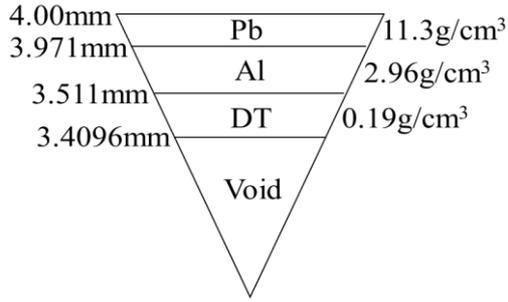

Fig. 3 An example fuel target structure in heavy ion inertial fusion (HIF).

the HIB pulse shape, the HIB pulse length and the HIB number density or current as well as the beam axis.

In a HIF reactor design, the reactor gas density should be about several torr~several tens torr in order to compensate the HIBs space charge neutralization at the final beam transport stage near the fuel target in a reactor [1, 22] and also to stabilize the filamentation and two-stream instabilities [1, 23]. The reactor gas density would be high compared with that in the laser fusion reactor, in which laser beams should be transported. In the HIF reactor gas, the driver HIB particle energy loss is negligible [1, 19].

In the ICF target implosion, the requirement for the implosion uniformity is stringent, and the implosion non-uniformity must be less than a few % [1, 10, 11]. Therefore, it is essentially important to improve the fuel target implosion uniformity. In general, the target implosion non-uniformity is introduced by a driver beams' illumination non-uniformity, an imperfect target sphericity, a non-uniform target density, a target alignment error in a fusion reactor, et al. The target implosion should be robust against the implosion non-uniformities for the stable reactor operation.

An ICF fuel target is injected from the outside of an ICF reactor transversely or vertically to a reactor center. In this study, we focus on the target alignment in a fusion reactor, and the target is injected transversely from a reactor side to the reactor center as shown in Fig. 1.

In order to release a sufficient fusion energy output, the target alignment error, that is, the target spatial displacement from the reactor center, should be small. For example, in HIF, our studies have presented that the target alignment error ($dr$) of about100μm~120μm is tolerable to release the fusion energy stably, when the 32 heavy ion beams (HIBs) as the energy driver are employed [24]. On the other hand, in laser fusion the target alignment error would be less than ~20-200μm [5-8]. Figure 2 shows the example simulation results for the HIBs illumination non-uniformity versus $dr$ and for the target gain versus $dr$. In Fig. 2 the direct-drive DT fuel target is employed as shown in Fig. 3. When the target alignment error $dr$ becomes larger than approximately 100~120 μm, the fuel pellet gain $G$, that is defined by the fusion energy output / the input driver energy, falls rapidly [24]. The example results in Fig. 2 demonstrate that the target alignment error should be less than approximately 100~120 μm.

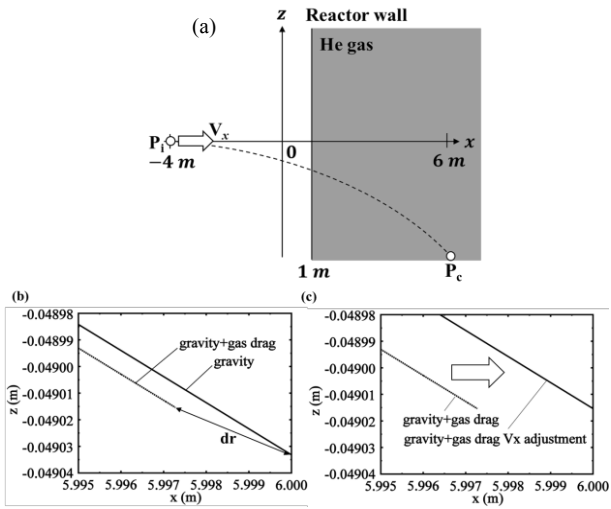

Fig. 4. (a) Schematic diagram for the fuel pellet injection into the reactor. (b) The fuel pellet orbit considering the gravity and/or the gas drag force, and (c) the fuel pellet orbit after adjusting the pellet longitudinal velocity $V_x$ to reach the chamber center at $x$=6 m.

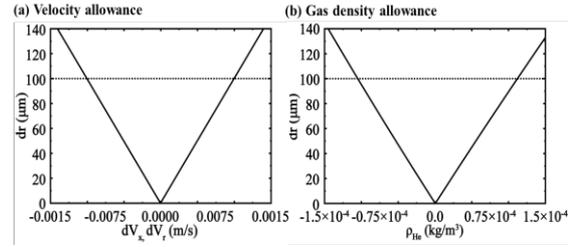

Fig. 5. (a)The longitudinal velocity $V_x$ or the vertical velocity $V_r$ allowance, and (b)the gas density allowance.

In this paper, the target injection speed is set to be 100m/s [14], and this target injection speed is typical in ICF. For example, when the reactor system is operated in ~10Hz, the interval between 2 shots must be less than 0.1sec, and the target ignition and burn induce a blast wave, which propagates in ~μsec [1] from the pellet position at the reactor center to the reactor wall. The target propagation time period is estimated by ~5m/(100m/s)~0.05s. As discussed in Sec. V, during the target traveling time inside of the reactor gas the Pb temperature of the cryo-target shown in Fig. 3 should be kept less than 7.2K to maintain the Pb superconductor. When the target injection speed is slower, it would be difficult to keep the target temperature lower. In this paper we also employ the typical target injection speed of ~100m/s [14].

In this paper, first we summarize the pellet injection concept used in this paper, and the results show that the target alignment is quite sensitive to the injection speed errors. In Sec. III, the magnetic correction method proposed in Ref. [14] is reviewed, and the permanent magnets are used to correct the target transverse speed error. In Sec. IV, we propose a new correction method to control the longitudinal target speed by using the electromagnets together with permanent magnets. The magnetic correction methods in Sec. III and IV, the cryo-target is assumed to be a superconductor. Therefore, in Sec. V we also check the temperature increase in the target Pb layer and also in the DT fuel layer. The paper presents that the magnetic correction method proposed in this paper is effective to construct a robust pellet injection system with a sufficiently small pellet alignment error.

## II. FUEL PELLET INJECTION

A HIF reactor chamber is filled with a chamber gas [1, 16]. The gas density should be high enough to compensate the focusing HIBs' space charge in a reactor chamber. This is a good way for the HIB neutralization transport in a fusion reactor chamber. In this section, we study the effect of the gas on transporting a fuel pellet to a fusion reactor center.

A fuel pellet is affected by the gravity and the gas drag force during the injection. The equation of motion is as follows:

$$m\frac{d^2\hat{r}}{dr^2} = -mg\hat{z} + \left(-6\pi\eta RV - \frac{1}{2}C_d\rho\pi R^2 V^2\right)\frac{\hat{V}}{V} \quad (1)$$

Here, $m$ is the pellet mass, $R$ is the pellet radius, $V$ is the pellet injection speed, $\eta$ is the gas viscosity, $C_d$ is the gas drag force coefficient and $Re$ is the Reynolds number. The coefficient of $C_d$ is obtained in Ref. [17] for the flow around a sphere:

$$C_d = \frac{24}{Re} + \frac{2.6\left(\frac{Re}{5.0}\right)}{\left(\frac{Re}{5.0}\right)^{1.52}} + \frac{0.411\left(\frac{Re}{2.63\times 10^5}\right)^{-7.94}}{1+\left(\frac{Re}{2.63\times 10^5}\right)^{-8.00}} + \frac{0.25\left(\frac{Re}{10^6}\right)}{1+\left(\frac{Re}{10^6}\right)} \quad (2)$$

When the reactor gas is He in a HIF reactor chamber, typically the He gas temperature would be 331 K, the mass density $6.61 \times 10^{-4}$ kg/m$^3$ and the viscosity $\eta$ is $2.53 \times 10^{-5}$ kg/m/s [1]. The speed of the spherical fuel pellet would be ~100 m/s [14], and in this study the pellet radius is 4 mm as shown in Fig. 3 [1]. The Reynolds number $Re$ is about 20.9 in this typical case. Therefore, if the fuel pellet is injected into the reactor chamber, the fuel pellet moves without oscillating in the fusion reactor chamber [25].

We examine the effect of the gravity and the gas drag force on the pellet trajectory. The simulation conditions are as follows: the fuel pellet is injected with 100 m/s along the $x$ axis from $x$=-4 m to $x$=6 m, that is, the reactor chamber center. The pellet radius is 4 mm and its mass is 0.286 g. The fusion reactor wall exists at $x$=1 m. The He gas is filled between $x$=1 m to 6 m. The simulation results are shown in Fig. 4(a). We measure the pellet position at time $t$=0.1 s, when the pellet reaches the reactor chamber center. The solid line indicates the pellet trajectory near the reactor center around $x$=6 m, including the gravity term. The dotted line shows the result including the gravity and the gas drag force. The fusion pellet drops slightly in the $-z$ direction by the gravity. At $t$=0.1 s, the gas drag force causes a significant pellet alignment error, that is, the spatial deviation of $dr$ ~3.269 mm. The allowable pellet alignment error is ~100-120μm [1, 24]. Accordingly, the deviation by the gravity and the gas drag force is too large. Therefore, the longitudinal pellet velocity should be adjusted to let the pellet reach the reactor chamber center at $t$=0.1 s as presented in Fig. 4(b) (the solid line). The adjusted pellet injection speed was 100.02730 m/s in this case.

We also investigate the velocity $V_x$ allowance and the gas density allowance considering the gravity and the gas drag force. The fuel pellet is injected from the point of $P_i$ = (-4 m, 0, 7.841 × 10$^{-3}$ m). The fusion reactor chamber center is set to the point of $P_c$ = (6 m, 0, -4.11743 × 10$^{-2}$ m) in this case. The fuel pellet velocity of $V_x$ is set to be 100.0273 m/s to compensate the gas drag force effect on the pellet alignment as shown above in Fig. 4(b). The fusion chamber gas pressure $P_{He}$ is 430 Pa. The alignment error $dr$ is the spatial deviation of the fuel pellet arrival point and reactor center at $t$=0.1 s. The simulation results in Fig. 5(a) shows the longitudinal velocity $V_x$ or the vertical velocity $V_r$ allowance. Figure 5(b) shows the gas density allowance to keep the fuel pellet alignment error smaller than ~100 $\mu$m. The gas density allowance window is sufficiently wide -15.9% to 16.8%. However, the longitudinal velocity or the vertical velocity allowance is strict, that is, ~-0.001% to ~0.001%. In this simulation result, we found that it is necessary to control the pellet injection velocity precisely.

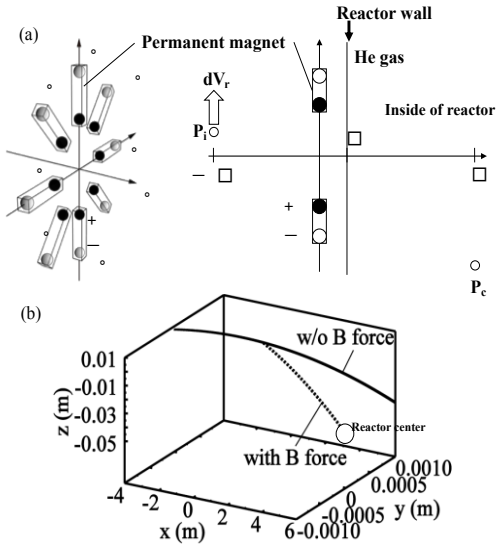

Fig. 6. (a)The fuel pellet injection with the magnetic correction system, in which eight permanent magnets are placed at $x$= 0, and (b) the fuel pellet orbits with the magnetic field correction and without the magnetic field correction at $dV_r$ (=$dV_y$) = 0.01 m/s.

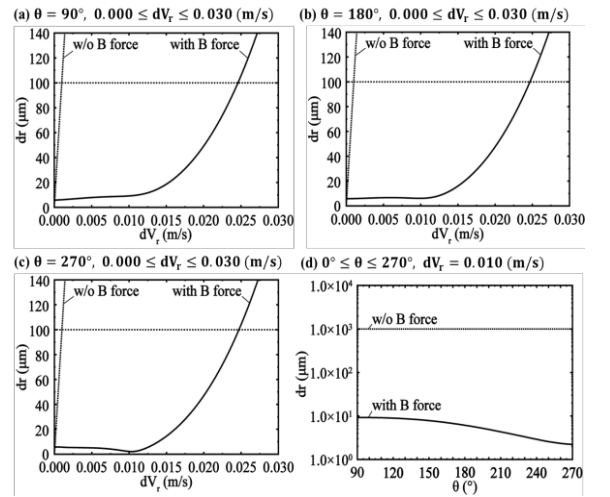

Fig. 7. The vertical velocity error allowances at (a)90, (b)180 and (c)270 degrees. (d)The target alignment error $dr$ at the vertical velocity error of $dV_r$ =0.01 m/s in the range of 90 to 270 degrees. The solid line indicates the results with the magnetic correction and the

## III. PELLET ALIGNMENT ERROR CORRECTION BY MAGNETS

In Sec. II, we found that a precise control is required in the pellet longitudinal and transverse injection velocity. In this section, we propose the magnetic field correction in order to make the target injection system robust [14]. The fuel pellet structure is shown in Fig. 3, and the outer surface is covered by Pb. The fuel pellet is cooled down by a liquid He. The cryo pellet coated by Pb becomes a superconductor, when its temperature is below 7.2 K [26]. The permanent magnets repel the superconductor due to the Meissner effect. We employ this effect in order to correct the fuel pellet injection velocity. The equation of motion considering magnetic field is as follows [14]:

$$\mathrm{m}\frac{d^2\hat{r}}{dr^2} = -mg\hat{z} + \left(-6\pi\eta RV - \frac{1}{2}C_d\rho\pi R^2 V^2\right)\frac{\hat{V}}{V} + F_{mag} \quad (3)$$

As an example, the equation of the magnetic force $F_{mag}$ by the magnet at $\theta=0$ degree is given below (see Fig. 6(a)):

$$F_{mag} = \frac{q^2 R\hat{r}(x, -(a-y), z)}{4\pi\mu_0}\left\{\frac{1}{(x^2+(a-y)^2+z^2-R^2)} + \frac{1}{R^2}\left(\frac{1}{x^2+(a-y)^2+z^2} - \frac{1}{(x^2+(a-y)^2+z^2-R^2)}\right)\right\} \quad (4)$$

The magnetic correction system is shown in Fig. 6 schematically. The fuel pellet and gas parameters are same as presented in Sec. II. In Fig. 6(a), the reactor wall is located at $x=1$m, the reactor center is at $x=6$m, and the eight permanent magnets are placed at $x=0$m. The magnetic charge $q$ is $3.54020 \times 10^{-5}$ Wb in this example case. The fuel pellet initial vertical velocity error $dV_r$ in the transverse direction is set to be 0.00 to 0.03 m/s in the directional range of 90 to 270 degrees (see Fig. 6(a)). **The fuel pellet example orbits are shown in Fig. 6(b) with the magnetic field correction and without the magnetic field correction at $dV_r (=dV_y) = 0.01$ m/s. The fuel pellet alignment error should be less than ~100 μm. The simulation results are shown in Figs. 7. In Figs. 7(a) to (c), the initial transverse velocity errors $dV_r = 0.00$ to 0.03 m/s are imposed at 90, 180 and 270 degrees. In Figs. 7(a)-(c), the vertical velocity errors' $dV_r$ allowances are shown by the dotted lines, when the magnetic correction is not employed.

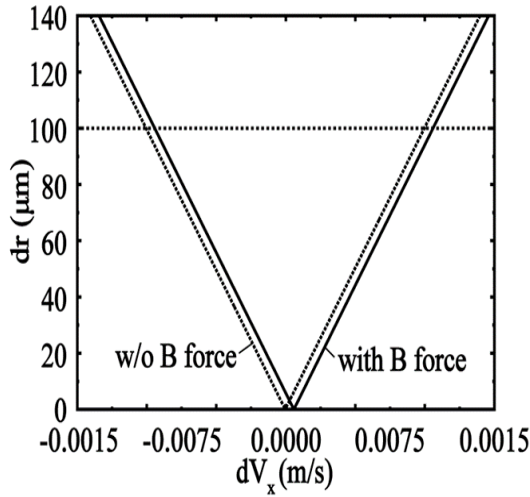

Fig. 8. The longitudinal velocity error $dV_x$ allowance. The solid line indicates the results with the magnetic field correction and the dotted line shows the results without the magnetic field correction. The longitudinal velocity error allowance is almost unchanged. In this figure just the permanent magnets are used to correct $dV_x$.

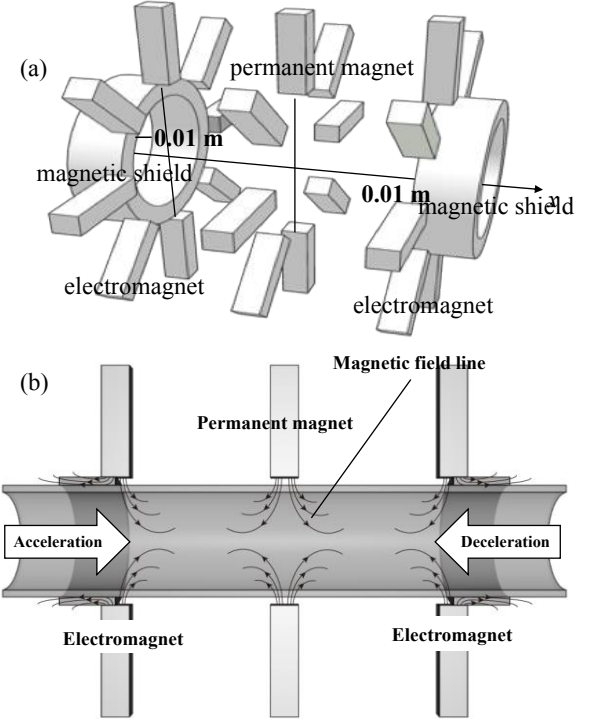

Fig. 9. (a)The longitudinal velocity $V_x$ control electromagnets and magnetic shields, and (b) the schematic magnetic fields.

However, when the magnetic correction is used, $dV_r$ allowances becomes larger ~0.0245 m/s (see solid lines in Fig. 7(a)-(c)). The allowable window for $dV_r$ is widened by ~24 times. Figure 7(d) shows the pellet alignment error of $dr$ versus $\theta$, at the vertical velocity error $dV_r = 0.01$ m/s. The results presented in Figs. 7 demonstrate that the magnetic correction system is quite effective to reduce the pellet alignment error against the transverse velocity error $dV_r$. However, the magnetic correction system in Fig. 6 is not effective for the longitudinal velocity error $dV_x$. In Fig. 8 the longitudinal velocity error allowance is shown, and it is found that the allowable window for the longitudinal velocity error $dV_x$ is kept to be almost unchanged. These eight permanent magnets cannot control the longitudinal velocity error $dV_x$. Therefore, we need a new method to make the pellet injection system robust against $dV_x$.

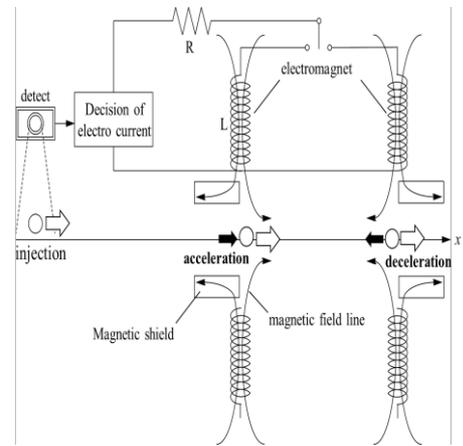

Fig. 10. Schematic diagram for the longitudinal velocity control system.

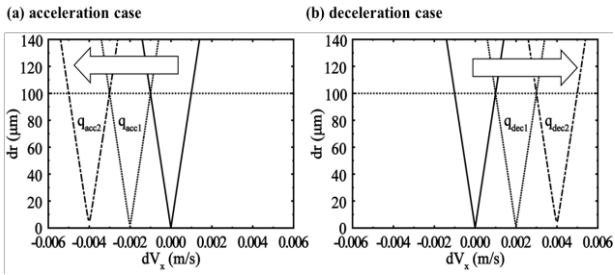

Fig. 11. (a) When the pellet longitudinal speed is slower than the target speed, the pellet is accelerated. The longitudinal velocity $V_x$ acceleration case. The longitudinal velocity error $dV_x$ allowance window is widened by the electromagnet and magnetic shielding system in Figs. 9 and 10. (b) The pellet speed is faster than expected, the pellet is decelerated.

## IV. LONGITUDINAL VELOCITY CONTROL

In Section III, we found that the magnetic correction system in Fig. 6 is robust against the transverse pellet injection velocity error of $dV_r$ but not for $dV_x$. In this section, we propose the control method for adjusting the longitudinal velocity to reduce the pellet alignment error. The longitudinal velocity control magnets model are shown in Fig. 9. A couple of electromagnets and magnetic shields [27] are placed in front and behind of the permanent magnets, which are used to correct the vertical velocity error. Figure 10 presents the schematic diagram for the longitudinal velocity control system. The fuel pellet longitudinal velocity is detected, just after the fuel pellet is ejected by a pellet ejection gun or so. The electric current is decided based on the detected velocity error $dV_x$. If the detected velocity error $dV_x$ is small beyond the allowance limit of the longitudinal velocity error of -0.001 m/s as shown in Fig. 5(a), the electromagnets in front of the permanent magnets are switched on. A part of magnetic field lines is shielded by the magnetic shields and just the magnetic force in the acceleration direction is applied on the fuel pellet (see Fig. 10). As a result, the fuel pellet is accelerated to fulfill the allowable window. Similarly, if the detected longitudinal velocity error $dV_x$ is larger than the longitudinal velocity allowance maximum +0.001 m/s, the electromagnets behind of the permanent magnets are switched on. A part of magnetic field lines is also shielded by the magnetic shields and only the magnetic force in the deceleration direction is applied to the fuel pellet. As a result, the fuel pellet is decelerated to reduce the pellet alignment error at the reactor chamber center.

We also estimate whether the electromagnets can be turned on before the fuel pellet reaches the acceleration and deceleration magnet system. In our simulation condition, the system is placed around $x$=0 m from the pellet ejection point at $x$=-4 m. The fuel pellet is injected around 100 m/s. Therefore, the fuel pellet reaches the system at 0.04 s after the pellet ejection. In this estimation, we think about a simple $LR$ circuit. The time constant $\tau$ is expressed $L/R$ and the electric current becomes in a steady state at about $5\tau$. Therefore, the relation of $5\tau$ < 0.04 s should be satisfied. We estimate $L$ and $R$ according to the system. The inductance $L$ would be approximately ~several µH for a small-scale coil, for example, with a length of ~1cm, a diameter of ~1cm and ~10 turns. The resistance $R$ would be ~$10^{-2}$V/A. So $5\tau$~$5L/R$~$10^{-3}$s. This estimation result demonstrates that the magnetic correction system would work under the simple detection and electromagnetic system with the magnetic shields. In an actual reactor system, we need to study the fast detection and circuit ignition system further in the near future. The whole magnetic correction system is outside of the reactor chamber as explained above and shown in Fig. 6.

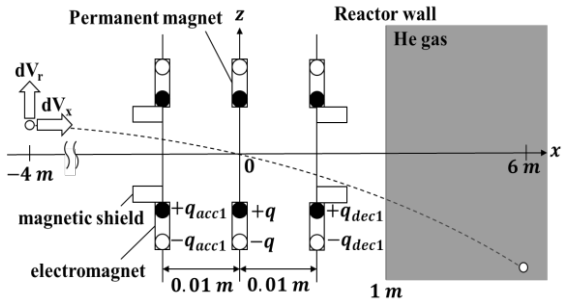

Fig. 12. Integrated pellet velocity error correction. The combination of the acceleration / deceleration electromagnets and the permanent magnets works well to reduce the transverse and longitudinal pellet alignment errors.

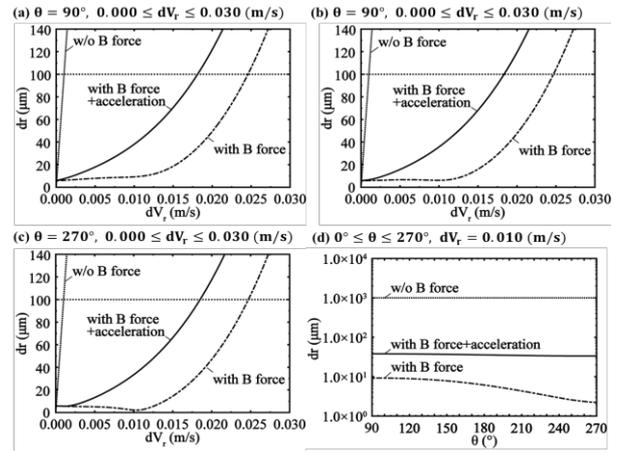

Fig. 13. The vertical velocity error $dV_r$ allowance at $\theta$= (a)90, (b)180 and (c)270 degrees. (d)The target alignment error $dr$ at the vertical velocity error $dV_r$ =0.01 m/s in the range of $\theta$=90 to 270 degrees in the combination of the acceleration electromagnets and permanent magnets.

We applied the magnetic correction system in Figs. 9 and 10 to simulate the longitudinal velocity control in the pellet acceleration and deceleration cases. The acceleration case is shown in Fig. 11(a). If the detected longitudinal velocity error $dV_x$ is around -0.002 m/s, the electromagnets placed in front of the permanent magnets are switched on and become $q_{acc1}$= 9.4222 × $10^{-6}$ Wb by changing the electric circuit current in Fig. 10. In this case, the fuel pellet is accelerated, and the longitudinal velocity error $dV_x$ allowance window is shifted to the next window nearby $dV_x$ ~-0.002 m/s. Similarly, if the detected longitudinal velocity error $dV_x$ is ~-0.004 m/s, the electromagnets are switched on and become $q_{acc2}$= 1.33233 × $10^{-5}$ Wb. The fuel pellet is accelerated and the longitudinal velocity error $dV_x$ allowance window is shifted to the next window nearby $dV_x$ ~-0.004 m/s. The deceleration case is shown in Fig. 11(b). If the detected longitudinal velocity error $dV_x$ is ~+0.002 m/s, the electromagnets placed behind the permanent magnets are switched on and become $q_{decc1}$ ~9.42234 × $10^{-6}$ Wb. The fuel pellet is decelerated and the longitudinal velocity error $dV_x$ allowance window is shifted to the next window nearby $dV_x$ ~+0.002 m/s. Similarly, if the detected longitudinal velocity error $dV_x$ is ~+0.004 m/s, the electromagnets are switched on and become $q_{dec2}$ ~1.33272 × $10^{-5}$ Wb. The fuel pellet is decelerated and the longitudinal velocity error $dV_x$ allowance window is shifted to the next window nearby $dV_x$ ~+0.004 m/s. By repeating the above process, the longitudinal velocity error allowance can be controlled and becomes wider by using the acceleration and deceleration system proposed in this paper in Figs 9 and 10.

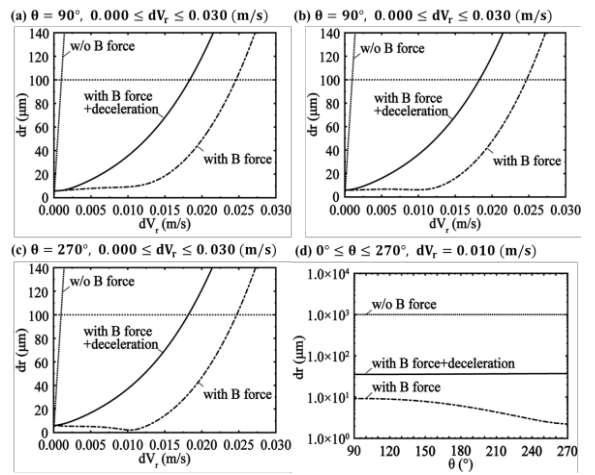

Fig. 14. The vertical velocity error $dV_r$ allowance in the (a)90, (b)180 and (c)270 degrees and (d)The target alignment error $dr$ at the vertical velocity error $dV_r$ 0.01 m/s in the range of 90 to 270 degrees in the combination of the deceleration electromagnets and permanent magnets.

Finally, we also check the integrated correction effect on the pellet alignment error including all the magnetic fields by the permanent magnet and the electromagnets shown in Fig. 12. In this case the pellet trajectory is designed to pass through the coordinate origin, so that the magnetic correction system works correctly. The integrated results of the pellet velocity error correction by the electromagnets and the permanent magnets are shown in Figs. 13(a) to (c). The vertical velocity errors $dV_r$ =0.00 to 0.03 m/s are given for $\theta$= 90, 180 and 270 degrees. In $\theta$= 90 degrees, the vertical velocity error $dV_r$ allowance becomes ~0.0181 m/s. It is 18.1 times larger than that without the magnetic force correction. In $\theta$=180 and 270 degrees, the vertical velocity error $dV_r$ allowance is ~0.0183 m/s and ~0.0185 m/s, respectively. Figure 13(d) shows the pellet alignment error $dr$ for the vertical velocity error of $dV_r$ =0.01 m/s in the range of $\theta$=90 to 270 degrees. The pellet alignment error $dr$ becomes ~33-38 $\mu$m. It is 28.6 times smaller than that without the magnetic field correction. The integrated results for the deceleration cases are shown in Figs. 14(a) to (c). In Fig. 14(d) the vertical velocity error $dV_r$ =0.01 m/s is fixed. The results presented here demonstrate that the magnetic correction system for the longitudinal and transverse velocity errors works very effectively to reduce the pellet alignment error in an inertial fusion reactor.

## V. FUEL PELLET TEMPERATURE CHANGE

In this section, we investigate the fuel pellet temperature change in the injection in the fusion reactor chamber. The fuel pellet is injected into the reactor gas, and its temperature would rise due to the interaction with the reactor gas. The issues are whether the solid DT of the fuel pellet is melted and whether the Pb layer of the fuel pellet is transferred from the superconductor to the normal conductor by the pellet temperature increase.

We also compute the fuel pellet temperature change in the reactor gas. The gas is He. The gas density $\rho_{He}$ is 6.61 × $10^{-4}$ kg/$m^3$, viscosity is $\eta$ =2.53 × $10^{-5}$ kg/m/s and the pressure $P_{He}$ is 430 Pa [1]. The fuel pellet initial temperature is 4.2 K. The fuel pellet is injected in the gas region at the velocity of 100 m/s. The simulation time is from 0.00 s to 0.05 s. The simulation results are shown in Figs. 15. Figure 15(a) shows the simulation spatial range. Figures 15(b)-(d) present the fuel pellet radial temperature profiles at $\phi$=0, 45, 90, 135 and 180 degrees at $t$=0.02, 0.04 and 0.05 s. The DT region is in 0.003496-0.003511 m, the Al region in 0.003511-0.003971 m and the Pb region in 0.003971-0.004000 m as shown in Fig. 3. At the 0.05 s, the highest temperature of DT region is 7.1 K at 0 degree. The DT should be less than the DT triple point of 19.79 K [28]. In this simulation case, the DT temperature is kept to be less than the triple point of DT. Therefore, the DT fuel is not melted. The highest temperature of the Pb layer is 7.3 K at 0 degree. The Pb layer of the fuel pellet is kept to be superconductor, when its temperature is below 7.2 K. The temperature of just the Pb surface becomes slightly beyond 7.2 K at 0.05s. As shown in Figs. 16, the temperature of Pb is kept below 7.2 K until $t$=0.0478 s. In this estimation, we assumed that the reactor gas would come back to the pellet injection port outside the reactor wall. In general, we would expect that the reactor gas does not reach the magnets' part. However, our estimation results suggest that we may need a gas backflow protection mechanism may be required [1].

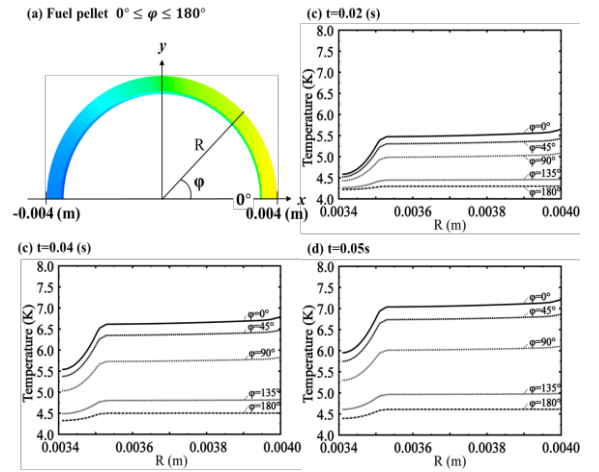

Fig. 15. (a)The simulation region. The temperature change of the fuel pellet at (b)0.0200, (c)0.0400 and (d)0.05 s in each degree.

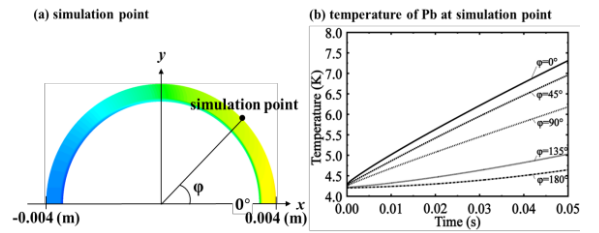

Fig. 16. (a)The simulation point. The temperature change of the Pb of the fuel pellet in each degree.

## VI. CONCLUSIONS

The allowable fuel pellet alignment error $dr$ is approximately 100~120 $\mu$m in order to obtain a sufficient fuel pellet gain $G$ [1, 24]. For the strict requirement for the pellet alignment in the fusion reactor, it was found that the allowances for the longitudinal velocity error $dV_x$ and the vertical velocity error $dV_r$ are $\pm 0.001$ m/s and stringent. We also found that the gravity and the reactor gas drag force are significant in the fuel pellet injection into the fusion reactor. The accurate control of the longitudinal and vertical pellet injection is required. We confirmed that the magnetic field correction by the eight permanent magnets proposed in Ref. [14] is effective to reduce the pellet transverse displacement error. However, the longitudinal velocity error $dV_x$ cannot be controlled by the permanent magnets in Fig. 6(a) [14]. So we proposed the new control system for the longitudinal velocity error $dV_x$. In the proposed system, the pellet longitudinal velocity error $dV_x$ is first detected, and $dV_x$ is reduced to fulfil the allowable requirement of $\pm 0.001$ m/s by the electromagnets and magnetic shields. The longitudinal velocity error $dV_x$ allowance window is widened by the acceleration / deceleration electromagnets and magnetic shielding system. The magnetic control system proposed in this paper would provide a new robust way to control the pellet alignment in the fusion reactor.

**Takeaki Kubo** received the M.S. degree from Utsunomiya University, Utsunomiya, Japan, in 2018.
He is currently with the Graduate school of Engineering, Utsunomiya University, Utsunomiya, Japan. He has worked on heavy ion nuclear fusion.

**Takahiro Karino** received the M.S. degree from Utsunomiya University, Utsunomiya, Japan, in 2016.
He is currently with the Ph.D., Graduate school of Engineering, Utsunomiya University, Utsunomiya, Japan. He has worked on heavy ion nuclear fusion.

**Hiroki Kato** received the M.S. degree from Utsunomiya University, Utsunomiya, Japan, in 2018.
He is currently with the Graduate school of Engineering, Utsunomiya University, Utsunomiya, Japan. He has worked on heavy ion nuclear fusion.

**Shigeo Kawata** (M'94) received the M.S. and Ph.D. degrees from the Graduate School of Science and Engineering, Tokyo Institute of Technology (TIT), Tokyo, Japan, in 1980 and 1985, respectively.

In 1981, he was a Research Associate with TIT, and in 1986, he was an Associate Professor with Nagaoka University, Nagaoka, Japan. He has been a Professor with the Utsunomiya University, Utsunomiya, Japan, since 1999. He has worked on laser particle acceleration, heavy ion nuclear fusion, and problem-solving environment in computational science and engineering.